\documentclass{aa}
\usepackage{txfonts}
\usepackage{graphicx}

\begin{document}

\title{Clustering environment of BL Lac object RGB 1745+398}

\author{H. Lietzen \inst{1} 
  \and K. Nilsson \inst{1} 
  \and L. O. Takalo \inst{1}
  \and P. Hein\"am\"aki\inst{1}
  \and P. Nurmi \inst{1}
  \and P. Kein\"anen \inst{1}
  \and S. Wagner \inst{2}}
\institute{Tuorla Observatory, University of Turku, 
V\"{a}is\"{a}l\"{a}ntie 20, FI-21500 Piikki\"{o}, Finland
\and Landessternwarte, K\"onigstuhl, Heidelberg, Germany}

\date{Received / Accepted }

\abstract
{}
{The BL Lac object RGB 1745+398 lies in an environment that makes it 
possible to study the cluster around it more deeply than the environments of 
other BL Lac objects. The cluster centered on the
BL Lac works as a strong gravitational lens, forming a large arc around 
itself. The aim of this paper is to study the environment and characteristics 
of this object more accurately than the environments of other BL Lac objects 
have been before.}
{We measured the redshifts of galaxies in the cluster from the absorption 
lines in their spectra. The velocity dispersion was then obtained from the 
redshifts. The gravitational lensing was used for measuring the mass at the 
center of the cluster. The mass of the whole cluster could then be estimated
using the softened isothermal sphere mass distribution. Finally, the richness 
of the cluster was determined by counting the number of galaxies near the BL Lac 
object and obtaining the galaxy-BL Lac spatial covariance function, $B_{gb}$.}
{The redshifts of nine galaxies in the field were measured to be near the 
redshift of the BL Lac object, confirming the presence of a cluster. 
The average redshift of the cluster is 0.268, 
and the velocity dispersion $\left(470^{+190}_{-110}\right)$ km s$^{-1}$. 
The mass of the cluster  
is $M_{500}=\left(4^{+3}_{-2}\right)\times10^{14}\,M_{\sun}$ 
which implies a rather massive cluster.
The richness measurement also suggests that this is a rich 
cluster: the result for covariance
function is $B_{gb}=\left(600\pm200\right)$ Mpc$^{1.77}$,
which corresponds to Abell richness class 1 and which is consistent with the 
mass and velocity dispersion of the cluster.}
{}

\maketitle

\section{Introduction}

The BL Lacertae objects are widely believed to be FR 1 radio galaxies viewed in 
the direction of the jet axis. There are several pieces of evidence for this
association (Urry \& Padovani \cite{Urry}):
the extended radio luminosity of BL Lac objects is similar to the flux of FR 1 
galaxies,
the host galaxies of BL Lacertae objects are large elliptical galaxies similar 
to FR 1 galaxies, 
and the luminosity function of the BL Lacs is what would be 
expected as Doppler-boosted luminosity function of FR 1s.

To find more evidence for this unification scheme, the environments of 
BL Lacs and FR 1s have to be studied. Since the properties of the cluster
environment do not depend on the viewing angle, the environments of both classes
should be similar.
Zirbel (\cite{Zirbel}) studied 123 groups and clusters of galaxies around
radio galaxies by determining a richness quantity $N^{-19}_{0.5}$,
which is obtained by counting galaxies that are brighter than $-19$ mag and
that surround the radio galaxy within a radius of 0.5 Mpc.
The results show that FR 1 galaxies are found in clusters
with average richness corresponding to Abell richness class 0
while FR 2 galaxies are even poorer.
According to Wold et al. (\cite{Wold}) the clusters with quasars also have 
an average richness class 0.

The cluster environments of BL Lac objects have not been studied as much as the 
environments of quasars. Wurtz et al. (\cite{Wurtz93}, \cite{Wurtz}) 
have done a statistical study of 50 objects, but only a few individual 
BL Lac environments have been studied more deeply. 
The results by Wurtz et al. (\cite{Wurtz93}, \cite{Wurtz}) 
show that most BL Lac host 
galaxies are in poor clusters; most of them have Abell richness class below 0.
This implies that BL Lac objects belong to poorer clusters than FR 1's, and
therefore they should not be similar objects.
Falomo et al. have studied a few individual objects:
H 0414+009 and PKS 2155-304 lie in moderately poor clusters with 
Abell richness class 0 
(Falomo et al. \cite{Falomo93a}, \cite{Falomo93b}), while
PKS 0548-322 belongs to a rich cluster with Abell richness class 2 
(Falomo et al. \cite{Falomo95}).

\begin{figure}
  \resizebox{\hsize}{!}{\includegraphics{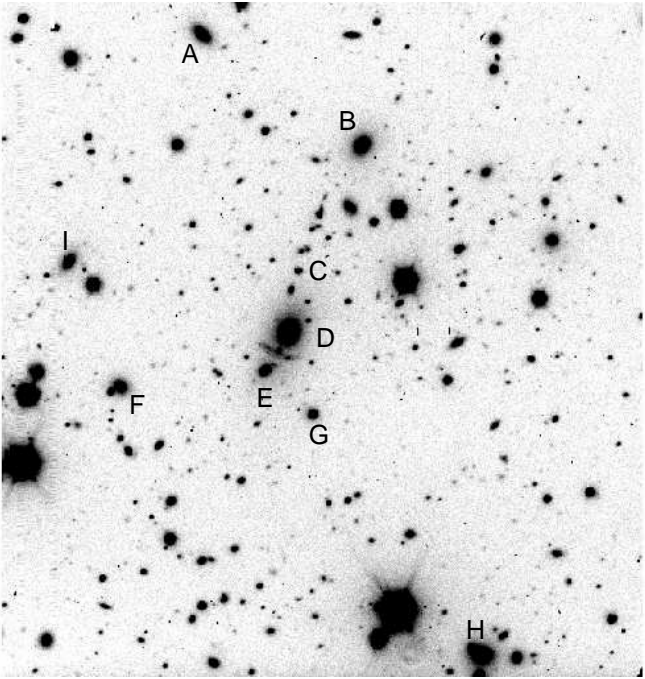}}
  \caption{ A 1800 s R-band exposure of RGB 1745+398 obtained with the NOT.
The field size is 207$\times$218 arcsec. Galaxy D is the central
brightest galaxy of the cluster harbouring the BL Lac nucleus. Objects
A-I are galaxies with measured radial velocities. The arc is visible
between galaxies D and E.}
  \label{arc}
\end{figure}

RGB 1745+398 has a significant advantage over most of the other known 
BL Lac objects:
the cluster works as a strong gravitational lens, forming a large blue arc
around itself (Fig. \ref{arc}; Nilsson et al. \cite{Nilsson}). 
Gravitational lensing gives us a possibility to observe also the dark matter, 
and thus to estimate the total mass of the cluster.

In gravitational lensing the light from a background object is bent by a large 
mass on its way. The result is a distorted image of the light source. Lensing 
makes it possible to observe light sources that would otherwise be too 
faint, but it also gives us information about the lens. Giant arcs are formed 
when the lens and the background source are nearly on the same line of 
sight. The arcs lie at the Einstein radius of the lensing clusters, and their 
radii depend on the mass inside the arc radius (Fort \& Mellier \cite{Fort}).

Besides RGB 1745+398, only one other BL Lac object with a possible 
gravitational arc 
has been found. Scarpa et al. (\cite{Scarpa}) report the discovery of almost 
perfect ring of radius $2.4''$ around BL Lac object 1517+656.
It was studied more closely by Beckmann et al. (\cite{Beckmann}) 
who measured the redshift and 
magnitude of the BL Lac, and estimated the mass and 
velocity dispersion of the cluster surrounding it.
The measured redshift of the object was $z=0.7$, but this result was uncertain
because the absorption they observed could have been caused also by a 
foreground object.
Unfortunately they did not
manage to measure the redshift of the arcs, which makes their mass and 
velocity dispersion estimates very uncertain.

RGB 1745+398 was part of the ROSAT-Green Bank sample of BL Lacertae objects
(Laurent-Muehleisen et al. \cite{Laurent}).
The radio source was identified with an elliptical galaxy 
at $z = 0.267$. 
The X-ray luminosity has been measured by Gliozzi et al. (\cite{Gliozzi}),
separately for the BL Lac object and the cluster. Their result for the 
BL Lac is $L_{\mbox{X,point}} = 2\times 10^{44}$ erg s$^{-1}$, and for the
cluster $L_{\mbox{X,ext}}=1.2\times 10^{44}$ erg s$^{-1}$, 
and their estimation for
the mass enclosed by the arc is 
$M_{\mbox{tot}}=1.6\times10^{12}-1.6\times10^{13}\,M_{\sun}$.
The blue arc lies 
8$''$ SE of the BL Lac object and has a length of 14$''$. Redshift of 
the arc is $z = 1.057$ (Nilsson et al. \cite{Nilsson}).

The target was also studied by Swinbank et al. (\cite{Swinbank}) who 
concentrated on the background galaxy which is seen as the arc. 
They modeled the lens cluster in order to form an undistorted 
image of the background 
galaxy. Their model used two galaxies near the arc: galaxies D and E in Fig.\ref{arc}.
In their best-fit model the velocity dispersion of the cluster is
560 km\,s$^{-1}$.

The purpose of this paper is to study the environment of RGB 1745+398. 
We measured the velocity dispersion from the redshifts of the cluster galaxies,
and estimated 
the mass $M_{500}$ using the softened isothermal sphere model.
The galaxy-BL Lac spatial covariance function, $B_{gb}$, was also obtained 
for determining the richness of the cluster. 
%The results were compared to the 
%results of Wurtz et al. (\cite{Wurtz}), which indicate that most clusters 
%with a BL Lac object are poor clusters.

The cosmological model used in this paper is the $\Lambda$CDM model with 
$H_0=71$ km s$^{-1}$Mpc$^{-1}$, $\Omega_m=0.3$ and $\Lambda=0.7$.

\section{Observations}

The observations were made at the 2.56 m Nordic Optical Telescope (NOT), 
La Palma, Canary Islands, in July 1997 and July 1998, 
using the ALFOSC instrument,
and in August 1999 at the Calar Alto 3.5 m telescope 
using the MOSCA instrument.
During the observations at the NOT there was some dust in the air 
and the seeing was 
mediocre, 1.0--1.5$''$ FWHM. At Calar Alto, seeing was between 
1.2 and 1.5 arcsec.
At NOT, $R$ and $B$ band images were obtained with 
exposure times of 900 s and field sizes of 6.5$\times$5.6 arcmin. 
The Calar Alto observations included $B$ band images with exposure time 900 s
and field size 11$\times$17 arcmin.
 The quality of the images at Calar Alto was worse than at the 
NOT, the largest problem being the lack of photometric standard stars, 
and we used them only to confirm the galaxy count results.
Images were 
bias-substracted and flat-fielded with twilight flats.
Photometric calibration was done using standard stars in the field of PG1633+099.

Longslit spectroscopy was obtained at the NOT using  
 a 1$''$ slit and
spectral range 5000--10250 \AA, with a resolution of 13.4 \AA.
The orientations of the slit were chosen in a way that covered as many 
possible galaxies near the BL Lac object as possible. There were nine spectra 
with 3600 s exposure in three different positions and two 1800 s exposures in 
two more positions.
The spectra were bias-subtracted using the overscan region and 
flat-fielded with a continuum lamp. They were wavelength-calibrated using 
calibration lamp exposures obtained before and after the science exposures 
and the calibration was tested by measuring the background 
sky emission lines before the background was subtracted. 
 The average errors in wavelength are approximately 0.5 \AA. 
This corresponds to 30 km/s at the object redshift.
The flux calibration 
was done using the standard star BD253941.

\section{Results}
\subsection{Mass estimation}

We observed spectra for nine galaxies, 
and their redshifts were obtained by 
measuring absorption lines $\lambda$4300 CH G, $\lambda$4861 H$_{\beta}$, 
$\lambda$5176 Mg b, $\lambda$5269 \ion{Fe}{i} and $\lambda$5892 NaD.
Line-of-sight velocities were calculated from the redshifts using the formula
\begin{equation}
  v_\parallel=\frac{cz-c\overline{z}}{1+\overline{z}},
\end{equation}
where $z$ is the redshift of the galaxy and $\overline{z}$ the average redshift
 of all the galaxies (Danese et al. \cite{Danese}).

\begin{figure}
  \resizebox{\hsize}{!}{\includegraphics{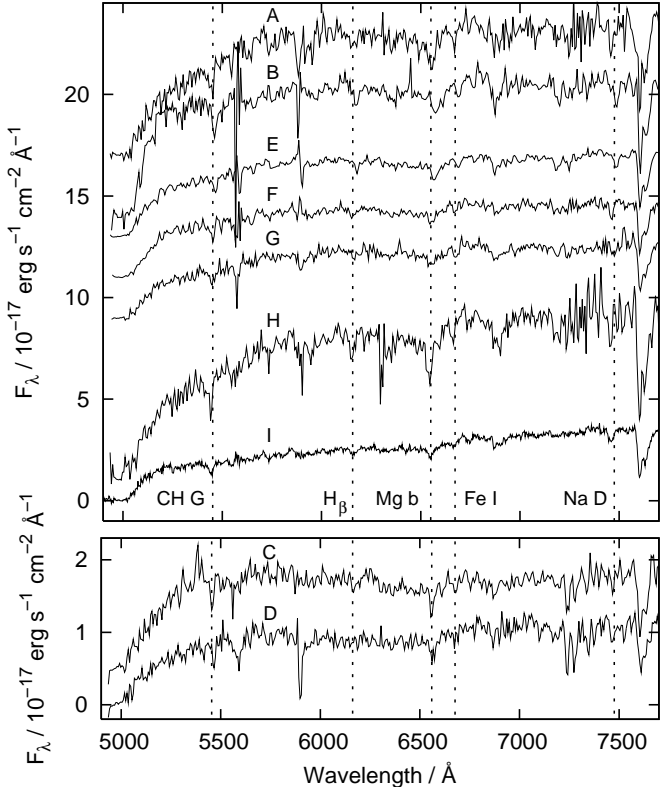}}
  \caption{ The spectra of galaxies A-I in Fig. 1. The vertical lines
indicate the positions of spectral features used to derive the radial
velocities. }
  \label{spectra}
\end{figure}

The locations of the measured galaxies can be seen 
in Fig. \ref{arc} and the spectra are shown in Fig \ref{spectra}.
The results for their redshifts and velocities and 
their measurement errors are 
presented in Table \ref{redshifts}. 
Galaxy D is the brightest galaxy containing the BL Lac nucleus.
The average redshift of the galaxies is
$z = 0.268\pm0.006$ which is well consistent with the BL Lac redshift 0.267 
measured 
by Laurent-Muehleisen et al. (\cite{Laurent}).
\begin{table}
  \caption{$R$ and $B$ magnitudes, redshifts, and velocities of 
    cluster galaxies. 
    Magnitudes were measured from the images obtained at the NOT}
  \label{redshifts}
  \centering
  \begin{tabular}{c c c c c c c}
    \hline\hline
    Object&$R$&$B$&$z$&$\Delta z$&$v_\parallel$&$\Delta v_\parallel$ \\
    \hline
    A &18.29 &20.98 &0.2659 & 0.0002 & -400 & 60 \\
    B &18.21 &20.81 &0.2705 & 0.0002 & +690 & 60 \\
    C &20.12 &23.51 &0.2680 & 0.0005 & +100 & 200 \\
    D &17.23 &18.85 &0.270  & 0.002  & +500 & 600 \\
    E &18.86 &21.44 &0.2701 & 0.0004 & +600 & 200 \\
    F &18.28 &20.42 &0.2666 & 0.0003 & -240 & 90 \\
    G &19.12 &21.70 &0.2671 & 0.0007 & -100 & 300 \\
    H &17.90 &20.58 &0.2650 & 0.0004 & -600 & 200 \\
    I &18.75 &21.29 &0.2656 & 0.0002 & -470 & 60 \\
    \hline
  \end{tabular}
  \end{table}

The radial velocity dispersion of the cluster is calculated using the methods 
of Danese et al. (\cite{Danese}), and the result is $\sigma_\parallel=
\left(470^{+190}_{-110}\right)$ km s$^{-1}$, measurement errors being 
68 \% confidence uncertainties. This result is consistent with the velocity
dispersion of Swinbank et al. (\cite{Swinbank}).

If we assume that the arc lies on the Einstein radius of the cluster the mass 
inside this radius is 
$M(8\arcsec)=(0.9\pm0.4)\times10^{13}M_{\sun}$. This confirms the result of
Nilsson et al. (\cite{Nilsson}).
The error in the Einstein radius was assumed to be 20 \%.
The $R$ band flux of the BL Lac host galaxy has been measured to be 
0.65 mJy, and so mass-to-luminosity ratio inside the $8''$ radius is 
$M/L_R\approx 38\,M_{\sun}/L_{\sun}$.

For estimating the total mass of the cluster, we assume the softened 
isothermal mass distribution with a density distribution
\begin{equation}
  \rho(r)=\frac{\sigma_{\parallel}^2}{2\pi G(r^2+r_c^2)}
\end{equation}
where $\sigma_{\parallel}$ is the radial velocity distribution and $r_c$ is 
the core radius.
Surface density at projected distance $\xi$ is 
\begin{equation}
  \Sigma(\xi)=\frac{\Sigma_0}{\sqrt{1+\xi^2/\xi_c^2}}\label{Sigma}
\end{equation}
where
$\Sigma_0=\frac{\sigma_{\parallel}^2}{2G\xi_c}$
is the surface density in the center of the cluster (Kormann et al. 
\cite{Kormann}).
This must be larger than the critical surface density of gravitational lensing 
$\Sigma_{cr}=\frac{c^2}{4\pi G}\frac{D_s}{D_dD_{ds}}$ where $D_s$, $D_d$ and 
$D_{ds}$ are distances to the source seen as the arc, to the lensing cluster 
and between the source and the cluster respectively. 
With these the maximum core 
radius can then be calculated, and the result is $\xi_c\approx10$ kpc. 
Surface density of the cluster center $\Sigma_0$
can be obtained by integrating equation \ref{Sigma} so that it gives an 
equation for the mass which is then set equal to the mass aquired using the 
lensing. Mass on any radius $r$ can then be calculated with equation
\begin{equation}
  M=\frac{c^2\xi_{arc}^2 D_s\left(\sqrt{1+\left(\frac{rc^2D_s}
{2\pi\sigma_{\parallel}^2D_dD_{ds}}\right)^2}-1\right)}
{4GD_dD_{ds}\left(\sqrt{1+\left(\frac{\xi_{arc}c^2D_s}
{2\pi\sigma_{\parallel}^2D_dD_{ds}}\right)^2}-1\right)}
\label{mass}
\end{equation}
where $\xi_{arc}$ is the distance of the arc from the cluster center 
in distance units. 

The cluster mass is calculated as the mass inside 
$r_{500}$, which is the distance
inside which the density is more than 500 times the critical density of the 
Universe, in this case about 1 Mpc.
The mass is $M_{500}=\left(4^{+3}_{-2}\right)\times10^{14}\,M_{\sun}$.
When comparing to other clusters, for example in RASS-SDSS survey 
(Popesso et al. \cite{Popesso}), we can see that this mass is similar to 
masses of other clusters with the same X-ray luminosity.
Bahcall et al. (\cite{Bahcall}) give relations between mass, 
velocity dispersion, luminosity and richness for clusters of galaxies. 
They use the mass inside an 0.6 Mpc radius from the cluster center. 
We calculated this mass using equation \ref{mass} with $r=600$ kpc, and 
the result is $\left(2.2^{+0.9}_{-0.6}\right)\times10^{14}\,M_{\sun}$.
The richness indicator $\Lambda_c$ related to this mass is between 60 and 90,
and there should be at least 30 galaxies in the cluster.
According to McNamara et al. (\cite{McNamara}), this refers approximately 
to Abell richness class 1 or 2. 
On average this relation between the mass and the richness 
is consistent with the theoretical models of the halo occupation distribution 
by Kravtsov et al. (\cite{Kravtsov}) and Zheng et al. (\cite{Zheng}).
According to Ledlow et al. (\cite{Ledlow}) also the X-ray 
luminosity indicates about the same richness although the relation between 
the richness and the luminosity is quite weak.
 On the whole, our results for the mass of the cluster points towards a 
relatively rich cluster, having Abell richness class \ $\sim 1$.

\subsection{Richness calculation}

For a direct measure of richness we obtained
the galaxy-BL Lac spatial covariance function amplitude, $B_{gb}$. 
The method used for this was initiated by Longair \& Seldner 
(\cite{Longair}), and it has been widely used for studying the environments
around different types of AGN (e.g. Wurtz et al. \cite{Wurtz}, Yee \& L\'{o}pez-Cruz \cite{Yee}, 
Wold et al. \cite{Wold}). We did this measurement 
using the NOT $R$ band image,
but also counted the number of the galaxies in the $B$ band images of 
both NOT and Calar Alto to
make sure the background count is not infected by the cluster because of the 
smaller field of view of ALFOSC.
%Unlike in other parts of this study, 
%in determination of $B_{gb}$ we used the low density cosmology model, with
%$\Omega_m=0.04$ and $H_0=50$ km s$^{-1}$ Mpc$^{-1}$ to 
%compare our results to Wurtz et al. (\cite{Wurtz}).

The determination of $B_{gb}$ was done in the following manner:

1. The sources which were closer than 500 kpc from the central galaxy of the 
cluster were analyzed using Source Extractor (Bertin \& Arnouts \cite{Bertin}) 
in order to separate the 
galaxies from the foreground stars. The radius 500 kpc at the distance of the 
cluster correspondes to $123''$ in angular units.

2. The galaxies in the area inside the 500 kpc radius, 
which were brighter than the limiting
magnitude, were counted. 

3. The number of background galaxies was determined by performing steps 
1 and 2 for a same-size area near the cluster. 

4. Angular cross-correlation function was calculated using the equation
\begin{equation}
  A_{gb}=\frac{N_{tot}-N_b}{N_b}\frac{3-\gamma}{2}\theta^{\gamma-1}
\end{equation}
where $N_{tot}$ is the number of galaxies within the circle of radius $\theta$ 
corresponding 500 kpc at the BL Lac redshift and $N_b$ is the number of 
background galaxies. The slope of the correlation function $\gamma$ was 
assumed to be 1.77.

5. The spatial covariance function amplitude was then obtained by scaling
$A_{gb}$ with universal luminosity function $\Phi(m_{lim},z)$. This is done by
using equation
\begin{equation}
  B_{gb}=\frac{N_gA_{gb}}{\Phi(m_{lim},z)I_{\gamma}}d_{\theta}^{\gamma-3}
\end{equation}
where $I_{\gamma}=3.78$ is an integration constant, $d_{\theta}$ the 
angular diameter distance to the cluster, $N_g$ the average surface density 
of galaxies, and it is obtained by integrating the universal luminosity
function over the distance. We use the luminosity function of Schechter 
(\cite{Schechter}) with the parameters $M^*=-20.9$ and $\alpha=-1.0$.
The $K$ and evolutionary corrections were taken from Poggianti (\cite{Poggianti}).

6. The measurement error for $B_{gb}$ was calculated according to
\begin{equation}
  \Delta B_{gb}/B_{gb}=\frac{\left[(N_{tot}-N_b)+1.3^2N_b\right]^{1/2}}
	 {N_{tot}-N_b}
\end{equation}
(Yee \& L\'{o}pez-Cruz \cite{Yee}).

\begin{figure}
  \resizebox{\hsize}{!}{\includegraphics{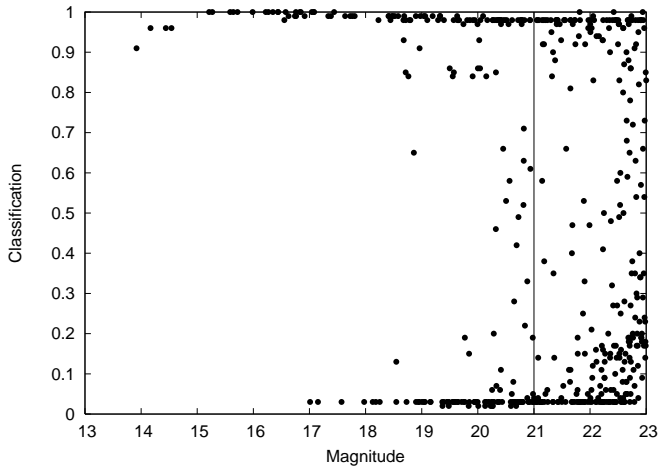}}
  \caption{Object classification for all the objects in the R band image. 
Classification value close to 1 means the object is a star, value close to 0 
points towards a galaxy. When the nature of the object cannot be determined 
the classification number gets a random value between 0 and 1.
Our limiting magnitude $R = 21$ is marked with the vertical line.}
  \label{class}
\end{figure}

We used limiting magnitude $R=21.0$ for determining $B_{gb}$. This corresponds 
to the absolute magnitude -19.7 at the redshift of the cluster. This is 
approximately the same as the limit used by Wurtz et al. (\cite{Wurtz}). As 
they explain, the background counts start rising faster above this magnitude, 
resulting with less accurate results. Another reason for not using a fainter 
limit is that the classification between stars and galaxies gets more uncertain 
with growing magnitude. This can be seen in Fig. \ref{class}, which shows that 
the number of unidentified objects increases at high magnitudes.

The background counts were determined using a ring-shaped area around the 
cluster with inner radius $123''$ and outer radius $174''$. This is very near 
the cluster, and it might be possible that outer parts of the cluster would 
affect the background. Because of this we complemented the galaxy counts also 
by the $B$ band images. In the NOT $B$ band image the background galaxies were 
counted at the same area as in the $R$ band image, while for the Calar Alto 
image we used a rectangular area starting at $200''$ to the East of the cluster 
center. The differences between the $B$ band images of these telescopes are 
small, there are from 1 to 
6 galaxies more in the NOT image background than in the Calar Alto image 
background at the same limiting magnitude. 
In the worst case this could cause the result for $B_{gb}$ to be 
400 Mpc$^{1.77}$ too small, which corresponds to an error of one in Abell 
richness class.

There is a good consistency between our background counts in the NOT images 
and the average background counts presented by Wold et al. (\cite{Wold}). For 
the $R$ band image with limiting magnitude the number of background galaxies on
our $\sim 50000$ square arcsecond counting area is 23 galaxies, which gives 
an average number of 6300 galaxies per square degree, while Figure 2 in 
Wold et al. (\cite{Wold}) shows this number is approximately 6000. 
This comparison implies that our background counts should not be strongly 
biased.

The number of galaxies inside the 500 kpc radius of 1745+398 at the limiting 
magnitude $R = 21$ is $N_{tot}=44$ galaxies, the background number being 
$N_b=23$. These values give $A_{gb}=0.0018$ rad$^{0.77}$, and 
$B_{gb}=595$ Mpc$^{1.77}$, with error of $\Delta B_{gb}/B_{gb}=0.37$.

Comparing our result to the results of Wurtz et al. (\cite{Wurtz}) is slightly 
problematic because they used a cosmology model with 
$H_0=50$ km s$^{-1}$Mpc$^{-1}$ and $q_0=0.02$. With this cosmology our result 
would be $B_{gb}=934$ Mpc$^{1.77}$. According to Yee \& Lopez-Cruz (\cite{Yee})
this corresponds to Abell richness class 1, and it is clearly richer than most 
clusters with a BL Lac object in Wurtz et al. (\cite{Wurtz}).

A simpler way to determine the richness of the cluster is to use $N_{0.5}$,
which is the excess of galaxies inside the 500 kpc radius of the cluster 
center. Bahcall (\cite{Bahcall81}) defines $N_{0.5}$ as the number of 
galaxies brighter than $m_3+2$, subtracted by the number of background galaxies
with the same limiting magnitude. The magnitude $m_3$ is the magnitude of the 
third brightest galaxy in the cluster, and in our case it is $R=18.2$ (target B
in Fig. \ref{arc}). The result is $N_{0.5}=13$, and it corresponds to Abell 
richness class 1.

In order to confirm the presence of a cluster at the correct redshift,
we also compared the NOT $R$ and $B$ band images, and calculated the $B-R$ 
colors of targets that were classified as galaxies in both fields. 
According to Fukugita et al. (\cite{Fukugita}) the $B-R$ color of a typical 
elliptical galaxy at redshift 0.2 is 2.39, and therefore we expect most of the 
galaxies at redshift 0.267 to have $B-R$ between 2 and 3. 
In order to get as large a number of galaxies as possible, we set the limiting 
magnitude as high as possible even if the target classification gets uncertain.
On $R$ band, targets could be found until magnitude $R = 23$, and on $B$ band 
until $B = 24$. With these limits we found 88 galaxies inside the 500 kpc 
radius and 75 at the background area in the
$R$ band image, and 89 galaxies inside the 500 kpc radius and 58 galaxies 
at the background area in the $B$ band image.
Of these galaxies 46 at the cluster area, and 35 at the background area were 
found and recognized as galaxies on both wavelength bands. 
Most of these galaxies were bright ones, with $R<21$. The color 
distribution of these galaxies is shown in table \ref{colors}.
\begin{table}
  \caption{Number of galaxies inside the 500 kpc radius 
($N_{cluster}$) and in the background area ($N_{background}$) at different 
colors. }
  \label{colors}
  \centering
  \begin{tabular}{c c c}
    \hline\hline
    $B-R$ color&$N_{cluster}$&$N_{background}$\\
    \hline
    $B-R<1$ & 4 & 4 \\
    $1<B-R<2$ & 18 & 17 \\
    $2<B-R<3$ & 22 & 13 \\
    $B-R>3$ & 2 & 1 \\
    \hline
  \end{tabular}
  \end{table}
Eight of the nine galaxies that we have a spectrum for have the $B-R$ color 
between 2 and 3; and as we can see in table \ref{colors}, this color bin is 
also where most of the excess galaxies at the cluster area are. Wake et al. 
(\cite{Wake}) have measured color-magnitude relations for 12 clusters at 
redshift $\sim 0.3$, and also the galaxies in those clusters have colors in this 
range more frequently than do the background galaxies.

\section{Conclusions}

We studied the cluster environment of RGB 1745+398 using images and 
spectra of the galaxies near the BL Lac. 
Our measurements confirm that there is a cluster of galaxies surrounding 
RGB 1745+398.
We measured the velocity dispersion and calculated the total mass of the 
cluster by assuming a softened isothermal sphere
for its mass distribution, and measured the richness of the cluster by 
determining the galaxy-BL Lac spatial covariance function amplitude $B_{gb}$.
This is the first time the cluster mass and dark matter content of a 
cluster with a
BL Lacertae object has been studied in such detail.

The velocity dispersion of the cluster is 470 km\,s$^{-1}$, which
is consistent with the result of Swinbank et al. (\cite{Swinbank}).
The mass, $4\times10^{14}\,M_{\sun}$ and the 
x-ray luminosity, $1.2\times10^{44}$ erg s$^{-1}$  
indicate that the cluster is a massive one. The richness of the cluster, 
$B_{gb}=600$ Mpc$^{1.77}$ is consistent with these results. We estimate the 
Abell richness class of the cluster is approximately 1.

%On the other hand, the $B_{gb}$
%value, 180 Mpc$^{1.77}$ indicates the cluster is a poor one, and should 
%therefore be also less massive.

%These results are unexpected because there is a correlation between richness 
%and the mass and the X-ray luminosity (Ledlow et al. \cite{Ledlow}). 
%The cluster seems to have more dark mass compared to the number of galaxies
%than most clusters.
%However, the correlation is not very 
%strong, and there are also other exceptions.
%An interesting question is whether this speciality of this cluster is caused 
%by the BL Lac. A comparison to other clusters with BL Lacs would give us an
%answer.

The other BL Lac cluster with a possible gravitational arc, 1517+656, 
could offer a
valuable reference point for comparing our results. 
Beckmann et al. (\cite{Beckmann}) estimate the lower limit for its mass 
inside the gravitational arc to be $2\times10^{12}\,M_{\sun}$. 
According to our estimation, this would mean the total mass of the whole 
cluster, inside a 1 Mpc radius, was 
$M>2\times10^{14}\,M_{\sun}$. This is close to our result for 
the cluster around 1745+398, but it is based on the assumption that the 
redshift of the arcs would be $z=2$, which is not based on measurements.
More measurements of this object could prove it really is a lensing system,
and give another accurate mass estimate.

The numerical simulations have predicted a model for AGN where matter merges 
into the supermassive black hole, and this causes the high luminosity (e. g. 
Hopkins et al. \cite{Hopkins}). 
Recent observations by Serber et al. (\cite{Serber}) propose that the local density 
excess of galaxies within 0.1--0.5 Mpc is likely to contribute to the triggering
of quasar activity through mergers and other interactions.
On the other hand, 
Zauderer et al. (\cite{Zauderer}) summarize many studies of the 
environments of different types of AGN, showing that there are not more than the 
average number of field galaxies surrounding most of the AGN. From this point 
of view, RGB 1745+398 is an object in an unusually rich environment.
More research is still needed for finding out how the AGN activity is related to
the environment on a larger scale.

This has been a detailed study of one cluster with a BL Lac object, but 
to get some answers for the unification schemes of BL Lac objects and
other AGN we should study more AGN in a similar fashion. Since gravitational 
arcs are so rare, most AGNs cannot be studied quite this accurately. 
One possibility is
to find out whether the amount of dark matter near AGN is different in different
types of objects. Estimates of mass from the velocity dispersion 
and the X-ray luminosities of the clusters
could be compared to the richness values. Also the large-scale surveys 
provide new possibilities for studying the environments of different kinds of 
AGN at different redshifts.

\begin{acknowledgements}
The authors would like to thank the anonymous referee for the useful comments 
and suggestions for improving this article.

Based on observations made with the Nordic Optical Telescope, operated 
on the island of La Palma jointly by Denmark, Finland, Iceland, Norway and 
Sweden, in the Spanish Observatorio del Roque de los Muchachos of the 
Instituto de Astrofisica de Canarias. 

The data presented here have been taken 
using ALFOSC, which is owned by
 the Instituto de Astrofisica de Andalucia (IAA) and operated at the Nordic 
Optical Telescope under agreement between IAA and the NBIfAFG of the 
Astronomical Observatory of Copenhagen.

Based on observations collected at the Centro Astron\'omico Hispano 
Alem\'an (CAHA) at Calar Alto, operated jointly by the Max-Planck 
Institut f\"ur Astronomie and the Instituto de Astrof\'isica de Andaluc\'ia 
(CSIC).
\end{acknowledgements}

\end{document}